\documentclass[preprint,amsmath,amssymb]{revtex4}
\usepackage{graphicx}
\usepackage{dcolumn}
\usepackage{bm}

\begin{document}
\title{The calculation of stress energy tensor from thermal properties of blackfold }
\author{  Z.~Amoozad $^{a}$ \footnote{Electronic
address:~z.amoozad@stu.umz.ac.ir }
and  J. ~Sadeghi $^a$ \footnote{Electronic address: ~pouriya@ipm.ir , corresponding author } }
\address{$^{a}$ Sciences Faculty, Department of Physics, University of Mazandaran, Babolsar, Iran \\
P.O.Box 47416-95447.}

\begin{abstract}
In this paper, we determine thermodynamical quantities for one kind of higher dimensional black holes. We take two charged and neutral blackfold which are higher dimension. We investigate the thermodynamical properties of such black hole at near horizon. By expanding those metrics around horizon, considering periodicity condition of imaginary time and applying some thermodynamical constraints, we extract stress energy tensor of blackfolds. We compare the thermodynamical properties of charged and neutral blackfold. We see here that the stress energy tensor of charged blackfold has brane-tension component in addition to thermal properties. Then by knowing that the spacetime of neutral blackfold is Ricci-flat and there is a nice relation between Ricci-flat and AdS spacetimes we determine the AdS form of Ricci-flat. Also we calculate the corresponding temperature which is satisfied by our previous calculation. Finally, we check the phase transition of blackfolds and show that there is not any critical point for them. So, the thermodynamical stability of both blackfolds will be proven, this result agree with Ref ~\cite{yy}. 
\end{abstract}
\maketitle

\section{Introduction}
\label{intro}

The investigation of stress energy tensor in hydrodynamics and fluid dynamical theories is very familiar and interesting for physicists. The role and appearance of this quantity in high energy physics is one of the most intriguing concepts which attracts much attentions. In curved spacetime and consequently in higher dimension paradigm, stress energy tensor is a local quantity (in contrast to the flat spacetime quantum field theory which has global physical quantities).

On the other hand in literature, it is nicely proved that Einstein's field equations could be written as a thermodynamical equation for the four dimensional spherically symmetric static and stationary spacetimes and also axisymmetric cases at the near horizon region. So the temperature, entropy, pressure and energy density of the black holes could be determined. This method shows elegantly that black hole's horizon is a thermodynamical system.

In high energy physics and in string theory the existence of spacetimes with dimensions higher than four is predicted. These spacetimes have new properties and require some new tools to work on. Using similar known techniques and definitions in 4-dimensional black holes, thermodynamical quantities of higher dimensional black holes have been determined by some nice methods~\cite{a,b,c,d,e}. Reviewing the methods discussed above, we extend the near horizon method to the blackfolds as new higher dimensional objects.

Extracting stress energy tensor by variation of action is the standard method which gives that tensor directly. At the other hand, the gravitational theory in d+1 dimensions will be corresponded to a statistical mechanics if the (d+1)'th dimension be periodic. After some computations, the temperature of the blackfold will be obtained. The entropy, pressure and energy density of blackfolds will be extracted. Also for the case of charged blackfold we calculate the charge and chemical potential. The thermal properties of such blackfolds help us to obtain the stress tensor of the charged and neutral blackfold. As we know the metric of the neutral blackfold is Ricci-flat and recently an interesting connection has been established between AdS and Ricci-flat spacetimes~\cite{p,q} and also the AdS form of the blackfold will be reproduced.
We note here that in the charged case the corresponding metric can not be transform to the AdS form because it is not Ricci-flat. So the stress energy tensor for the given metric is extracted just by the method of near horizon expansion.

The paper is organized as follows. In the next section some properties and metrics of charged and neutral blackfolds have been reviewed. In section 3  the near horizon expansion of charged blackfold will be present. Then by choosing variable and changing the metric as the form of Euclidian metric and applying periodicity condition of imaginary time, we have obtained temperature of the charged blackfold. By using thermodynamical relations we determine charge, chemical potential, pressure, entropy and energy density. At the end of this section we use the thermal properties and show the form of charged blackfol's stress energy tensor and discuses the general corresponding properties. In section 4 we repeat all the previous calculation for the neutral black fold. In addition to the above results, the AdS form of this Ricci-flat spacetime lead us to obtain temperature again, and explain the corresponding results. In section 5 we try to realize and check the phase transition of both cases\cite{ss} and show that the stability condition completely agree with ref~\cite{yy}. In the final section we conclude and note some points.

 \section{General properties of blackfolds}
\label{sec:1}
In high energy physics and string theory, introducing new dimensions (more than four) often solves some fundamental problems. Extensions of Schwarzschild black hole to higher dimensions by Kaluza-Klein theory and introducing one or more spatial extra dimensions is a simple example with interesting properties~\cite{g}. If the worldvolume of a black p-brane bent into the shape of a compact hypersurface, for instance that of a torus or a sphere, many new geometries and topologies of black hole horizon would obtain. As we know the effective theory of the corresponding worldvolume comes with following expressions: I- Worldvolume of black brane is not exactly flat, II- Worldvolume is not in stationary equilibrium, III- Deviation from the flat stationary black brane be on scales much longer than the brane thickness.
 Black branes whose worldvolume is bent into the shape of a submanifold of a background spacetime have been named blackfolds~\cite{r,s,t,u}.
Blackfolds have two length scales which are given by,
\begin{equation}
\label{eq:3}
\textit{l}_M \sim (GM)^{1/D-3}\;\;\;\;\  ,  \;\;\;\;\;\;\ \textit{l}_J \sim \frac{J}{M}.
\end{equation}
In case which $\textit{l}_J \gg \textit{l}_M $ , this separation suggests an effective description of long wavelength dynamics.

In supergravity and low energy limit of string theory, black branes have charges. One of the best choices is black p-branes that carry charges of Ramond-Ramond field strength type $F_{(p+2)}$. The charged dilatonic black p-brane solution of action in $D=n+p+3$ spacetime dimension is,
\begin{equation}
\label{eq:6}
I=\frac{1}{16\pi G}\int dx^{D} \sqrt{-g}(R-\frac{1}{2}(\partial \phi)^{2}-\frac{1}{2(p+2)!}e^{a\phi} F^{2}_{(p+2)}),
\end{equation}
where
\begin{equation}\label{eq:x}
a^{2}=\frac{4}{N}-\frac{2(p+1)n}{D-2}  \;\;\;\;\;\;  ,  \;\;\;\;\;\;    n=D-p-3.
\end{equation}
The flat black p-brane solution reads,

\begin{equation}
\label{eq:7}
ds^{2}=H^{\frac{-Nn}{D-2}}(-fdt^{2}+{\sum _{i=1}}^{p}{dz_{i}}^{2})+H^{\frac{N(p+1)}{D-2}}(f^{-1}dr^{2} +r^{2}{d\Omega ^{2}}_{n+1}),
\end{equation}

\begin{eqnarray}\label{eq:8}
\begin{split}
e ^{2\phi}&=H^{aN}   \;\;\;,\;\;\; A_{p+1}=\sqrt{N}\coth \alpha (H^{-1}-1)dt \wedge dz_{1} \wedge dz_{2} ... \wedge dz_{p}, \\
H&=1+\frac{{r_{0}}^{n}\sinh ^2\alpha}{r^{n}} \;\;\;\;\;\;\;\;\; ,  \;\;\;\;\;\;\;\;\;    f=1-\frac{{r_{0}}^{n}}{r^{n}}.\;\;\;\;\;\;\;\
\end{split}
\end{eqnarray}

where $\sinh \alpha$ is related to the boost of the blackfold which is a Lorentz transformation of some of directions of p-brane. As $a^{2}\geq0$,  the parameter $N$ is not arbitrary,
\begin{equation}
\label{eq:9}
N\leq2(\frac{1}{n}+\frac{1}{p+1}).
\end{equation}

In string/M theory, N is an integer up to 3 (when $p\geq1$)that corresponds to the number of different types of branes in an intersection.

As mentioned before, blackfolds can have $ \frac{\textit{l}_{M}}{\textit{l}_{J}}\rightarrow 0 $, so that a flat black brane could be produced. The effective theory describes the collective dynamics of a black p-brane, the geometry is given by,
\begin{equation}
\label{eq:4}
 {ds^{2}}_{p-brane}=-(1-\frac{{r_{0}}^{n}}{r^{n}})dt^{2}+\sum _{i=1}^{p}{dz_{i}}^{2}+{(1-\frac{{r_{0}}^{n}}{r^{n}})}^{-1}dr^{2} +r^{2}{d\Omega ^{2}}_{n+1},
\end{equation}
where $r_{0}$ is the thickness of the horizon.
Here $(p+1)$ coordinates are on the worldvolume of the blackfold and $(D-p-1)$ coordinates are transverse directions to the worldvolume.

For isotropic worldvolume theory, in case of lowest derivative order the stress tensor will be same as the isotropic perfect fluid, which is given by,
\begin{equation}
\label{eq:5}
T^{ab}=(\varepsilon+p)u^{a}u^{b}+p\gamma^{ab},
\end{equation}
where $\varepsilon $ and $p$ are the energy densityis and pressure respectively.

\section{Near horizon expansion of charged blackfolds}

In this section we work on thermal properties of charged blackfold and try to find it's stress energy tensor. For the case of charged dilatonic blackfold the metric is,
 \begin{equation}
\label{eq:28}
{ds^{2}}_{p-brane}=A{dt}^{2}+B{dz_{i}}^{2}+Cdr^{2} +F_{\alpha}{d\psi ^{2}}_{\alpha},
\end{equation}
and
\begin{equation}
\label{eq:29}
\begin{split}
A&=-(1+\frac{{r_{0}}^{n}\sinh ^{2}\alpha}{r^{n}})^{\frac{-Nn}{D-2}}(1-\frac{{r_{0}}^{n}}{r^{n}}),
\\
B&=(1+\frac{{r_{0}}^{n}\sinh ^{2}\alpha}{r^{n}})^{\frac{-Nn}{D-2}},
\\
C&=(1+\frac{{r_{0}}^{n}\sinh ^{2}\alpha}{r^{n}})^{\frac{N(p+1)}{D-2}}(1-\frac{{r_{0}}^{n}}{r^{n}})^{-1},
\\
F_{\alpha}&=F_{\alpha}(r,\psi_{\alpha}).
\end{split}
\end{equation}

 In this method we must expand the metric around horizon; $r\rightarrow r_{0}+u$ and $u\ll r_{0}$;
 \begin{equation}
\label{eq:12}
\lim_{r \rightarrow r_{0}+u} (1-\frac{{r_{0}}^{n}}{r^{n}})=\frac{nu}{r_{0}},
\end{equation}

 so the metric takes the following form,
 \begin{eqnarray}
 \label{eq:30}
ds^{2}=&-&(1+\sinh ^2\alpha)^{-\frac{n}{8}}\frac{nu}{r_{0}}dt^{2}+(1+\sinh ^2\alpha)^{-\frac{n}{8}}{dz_{i}}^{2}+
(1+\sinh ^2\alpha)^{\frac{8-n}{8}}\frac{r_{0}}{nu}du^{2} \nonumber \\
&+&(1+\sinh ^2\alpha)^{\frac{8-n}{8}}r_{0}^{2}{d\Omega ^{2}}_{n+1}.
\end{eqnarray}
If the $(d+1)$'th dimension be periodic, the Euclidian gravitational theory in $d+1$ dimensions will be equivalent to statistical mechanics in $d$ dimensions ~\cite{w}, so we take,
\begin{equation}
\label{eq:14}
dr^{2}\rightarrow du^{2},\;\;\;\;\;\;\; t=\tau \rightarrow -it,
\end{equation}
and by changing the variable as,
\begin{equation}
\label{eq:16}
\frac{du^{2}}{u}=d\rho ^{2}, \;\;\;\;\;\;\; \rightarrow\;\;\;\;\;\;\ \rho=2 \sqrt{u},
 \end{equation}
one can obtain,
 \begin{eqnarray}
 \label{eq:31}
ds^{2}&=&(1+\sinh ^2\alpha)^{\frac{8-n}{8}}\frac{r_{0}}{n}\left\{(1+\sinh ^2\alpha)^{-1}\frac{n^{2}\rho^{2}}{4r_{0}^{2}}d\tau^{2}+d\rho^{2}\right\}+(1+\sinh ^2\alpha)^{-\frac{n}{8}}{dz_{i}}^{2} \nonumber \\
 &+&(1+\sinh ^2\alpha)^{\frac{8-n}{8}}r_{0}^{2}{d\Omega ^{2}}_{n+1},
\end{eqnarray}

Now we consider the variable $\theta=(1+\sinh ^2\alpha)^{-\frac{1}{2}}\frac{n}{2r_{0}}\tau$, it must obey the periodicity condition,
\begin{equation}
 \label{eq:32}
 \tau\rightarrow\tau+\beta\;\;\;\Rightarrow\;\;\;\; (1+\sinh ^2\alpha)^{-\frac{1}{2}}\frac{n}{2r_{0}}\tau\rightarrow (1+\sinh ^2\alpha)^{-\frac{1}{2}}\frac{n}{2r_{0}}\tau+(1+\sinh ^2\alpha)^{-\frac{1}{2}}\frac{n}{2r_{0}}\beta,
 \end{equation}

\begin{equation}
 \label{eq:33}
 \theta\rightarrow \theta+2\pi \;\;\Rightarrow\;\;\;\; (1+\sinh ^2\alpha)^{-\frac{1}{2}}\frac{n}{2r_{0}}\tau\rightarrow (1+\sinh ^2\alpha)^{-\frac{1}{2}}\frac{n}{2r_{0}}\tau+ 2\pi , \;\;\;\;\;\;\;\;\;\;\;\;\;\;\;\;\;\;\;\;\;\;\;
 \end{equation}
so the temperature of blackfold takes the following form,
\begin{equation}
 \label{eq:34}
T=\frac{1}{\beta}=\frac{n}{4\pi r_{0}}(1+\sinh ^2\alpha)^{-\frac{1}{2}}.
 \end{equation}
This result is exactly the temperature which could be obtained by Hawking temperature relation.

We can  obtain the Bekenestein-Hawking identification between horizon area and entropy  by compactifying the $p$ directions along the brane with any given point on worldvolume. This help us to determine the entropy of blackfold. Thus the entropy density $s$ will be following form,

\begin{equation}
 \label{eq:35}
s=\frac{A}{4G}=\frac{(1+\sinh ^2\alpha)^{\frac{1}{2}}}{4G}r_{0}^{(n+1)}\Omega_{n+1},
 \end{equation}
 
 The chemical potential of charged blackfold could be obtained by the difference between potential at the horizon and the potential at the $r\rightarrow\infty$. So,
 
 \begin{equation}
 \label{eq:351}
\Phi_{p}=A_{(p+1)}\mid_{\infty}-A_{(p+1)}\mid_{r_{0}}=tanh\alpha,
 \end{equation}
 
 The electric charge would be considered as the measure of electric charge at infinity. As $F_{(p+2)}$ is the background strength field which couples with charge of blackfold we have,
 
  \begin{equation}
 \label{eq:352}
Q=\frac{1}{8\pi G (p+2)!}\oint e^{-2a\phi}F^{\mu_{1}...\mu_{p+2}}ds_{\mu_{1}...\mu_{p+2}},
 \end{equation}
by integration on hypersurface $S^{n+1}\times IR^{p}$ on $r\rightarrow\infty$ the density of electric charge will be,

\begin{equation}
 \label{eq:361}
Q =\frac{\Omega_{n+1}}{16\pi G}nr_{0}^{n}sinh\alpha cosh\alpha,
 \end{equation}

 Now we can determine $P$ as,
\begin{equation}
 \label{eq:36}
dP=sdT-\Phi dQ\;\;\;\; , \;\;\;\;\;\;\;\;\;P =-\frac{\Omega_{n+1}}{16\pi G}r_{0}^{n}(1+nsinh^{2}\alpha),
 \end{equation}
and by using $d\varepsilon=T ds+ \Phi dQ$ the energy density is,
 \begin{equation}
 \label{eq:37}
\varepsilon =\frac{\Omega_{n+1}}{16\pi G}r_{0}^{n}{1+n(1+sinh^{2}\alpha)},
 \end{equation}
Here also the eqs.(18-24) give us opportunity to obtain the corresponding stress energy tensor. So the stress energy tensor of chrged blackfold takes the following form
\begin{equation}
\label{eq:26}
T_{ab}=(\epsilon+P)u_{a}u_{b}+P\eta_{ab},
\end{equation}

 \begin{equation}
\label{eq:38}
T_{ab}=\frac{\Omega_{n+1}r_{0}^{n}}{16\pi G}(nu_{a}u_{b}-\eta_{ab})-\frac{\Omega_{n+1}r_{0}^{n}}{16\pi G}sinh^{2}\alpha\eta_{ab}.
\end{equation}

\begin{equation}
\label{eq:381}
T_{ab}=\frac{\Omega_{n+1}r_{0}^{n}}{16\pi G}(nu_{a}u_{b}-\eta_{ab})-\Phi_{p}Q_{p}\eta_{ab}.
\end{equation}
this result shows that in charged blackfold we have brane-tension component in addition to thermal properties.

\section{Near horizon expansion of neutral blackfold}
Now we are going to study the neutral blackfold and obtain the thermal properties. The thermal properties help us to arrange the stress energy tensor. In order to do such processes we consider appropriate metric which is given by,
\begin{equation}
\label{eq:11}
 {ds^{2}}_{p-brane}=-(1-\frac{{r_{0}}^{n}}{r^{n}})dt^{2}+\sum _{i=1}^{p}{dz_{i}}^{2}+{(1-\frac{{r_{0}}^{n}}{r^{n}})}^{-1}dr^{2}
 +r^{2}{d\Omega ^{2}}_{n+1}.
\end{equation}

 In this method we must expand the metric around the horizon, so by considering $r\rightarrow r_{0}+u $ and $u\ll r_{0}$, so the metric takes the following form,
\begin{equation}
\label{eq:13}
 {ds^{2}}_{p-brane}=-(\frac{nu}{r_{0}})dt^{2}+\sum _{i=1}^{p}{dz_{i}}^{2}+\frac{r_{0}}{nu}du^{2} +{r_{0}}^{2}{d\Omega ^{2}}_{n+1},
\end{equation}
as we have done in the previous section, the Euclidian spacetime could be given by,
\begin{equation}
\label{eq:14}
dr^{2}\rightarrow du^{2},\;\;\;\;\;\;\; t=\tau \rightarrow -it,
\end{equation}
so,
\begin{equation}
\label{eq:15}
 {ds^{2}}_{p-brane}=-(\frac{nu}{r_{0}})d\tau^{2}+\sum _{i=1}^{p}{dz_{i}}^{2}+\frac{r_{0}}{nu}du^{2} +{r_{0}}^{2}{d\Omega ^{2}}_{n+1},
\end{equation}
again by changing the variables,
\begin{equation}
\label{eq:16}
\frac{du^{2}}{u}=d\rho ^{2}, \;\;\;\;\;\;\; \rightarrow\;\;\;\;\;\;\ \rho=2 \sqrt{u},
 \end{equation}
 then
 \begin{equation}
\label{eq:17}
 {ds^{2}}_{p-brane}=\frac{r_{0}}{n}\{\frac{n^{2}\rho^{2}}{4{r_{0}}^{2}}d\tau^{2}+d\rho^{2}\} +\sum _{i=1}^{p}{dz_{i}}^{2}+{r_{0}}^{2}{d\Omega ^{2}}_{n+1},
\end{equation}
by taking $\theta=\frac{n\tau}{2r_{0}}$ we can rewrite,
\begin{equation}
\label{eq:18}
{ds^{2}}_{p-brane}=\frac{r_{0}}{n}\{\rho^{2}d\theta^{2}+d\rho^{2}\} +\sum _{i=1}^{p}{dz_{i}}^{2}+{r_{0}}^{2}{d\Omega ^{2}}_{n+1},
\end{equation}
as we can take the following,
\begin{equation}
\label{eq:19}
\tau\rightarrow\tau+\beta\;\;\;\;\;\;\;\Rightarrow\;\;\;\; \frac{n}{2r_{0}}\tau\rightarrow \frac{n}{2r_{0}}\tau+\frac{n}{2r_{0}}\beta,
\end{equation}
and
\begin{equation}
\label{eq:20}
\theta\rightarrow\theta+2\pi,
\end{equation}
so one can easily find
\begin{equation}
\label{eq:21}
\frac{n}{2r_{0}}\beta=2\pi \;\; , \;\;\;T=\frac{1}{\beta}\;\;\;\Rightarrow \;\;\;\;\; T=\frac{n}{4r_{0}\pi}.
\end{equation}
which is the temperature which can be extracted from Hawking temperature relation.
Now we can determine entropy by Bekenestein-Hawking identification as follows,
\begin{equation}
\label{eq:22}
s=\frac{A}{4G}=\frac{r_{0}^{n+1}\Omega _{n+1}}{4G}.
\end{equation}

After finding $s$ we can determine $P$ as,
 \begin{equation}
\label{eq:23}
P=-\frac{\Omega_{n+1}r_{0}^{n}}{16\pi G},
\end{equation}
and by using $d\varepsilon=T ds$ the energy density is,
\begin{equation}
\label{eq:24}
\varepsilon=\frac{\Omega_{n+1}}{16\pi G}r_{0}^{n},
\end{equation}
So obviously we prove that the following thermodynamic relations satisfied by the above thermodynamical quantity.
 \begin{equation}
\label{eq:25}
\varepsilon+P=Ts.
\end{equation}

then
\begin{equation}
\label{eq:27}
T_{ab}=\frac{\Omega_{n+1}r_{0}^{n}}{16\pi G}(nu_{a}u_{b}-\eta_{ab}).
\end{equation}
We have exactly found the stress tensor as we expected. This tensor is conserved and traceless. It can be obtained from an ADM-type prescription~\cite{x},or equivalently, from the Brown-York quasilocal stress-energy tensor~\cite{f}. the equation (\ref{eq:27}), based on general covariance for the intrinsic fluctuations, obeys the relation
\begin{equation}
\label{027}
D_{a}T^{ab}=0,
\end{equation}
 where $D_{a}$ is the covariant derivative for the worldvolume metric~\cite{s}. Since the extracted stress tensor could be regarded as the stress energy tensor of a perfect fluid, the above constraint is hydrodynamic Euler equation.

Now by using the method of AdS/Ricci-flat correspondence the AdS form of the neutral blackfold  can be reproduced~\cite{v},
\begin{equation}
\label{eq:x}
 {ds^{2}}_{\Lambda}=-\frac{1}{r^{2}}(1-\frac{{r}^{d}}{{r_{0}}^{d}})dt^{2}+ \frac{1}{r^{2}}({dz_{i}}^{2}+d \vec{y}^{2})+\frac{1}{r^{2}(1-\frac{{r}^{d}}{{r_{0}}^{d}})}dr^{2}.
\end{equation}
By expanding this metric near horizon and using the same method we will arrive at,
 \begin{equation}
 \label{eq:y}
 T=\frac{-d}{4\pi r_{0}},
 \end{equation}
We explain here when we transform from Ricci-flat to AdS we must change $n\rightarrow-d$.

Surprisingly we can see that although the entropy and temperature of charged blackfold differ from neutral blackfold, the concluded stress tensor takes the same form as the charged blackfold. This proves that expansion of charged and neutral blackfold near horizon takes the same local distribution of energy. It arises from effective description of blackfolds and compactifing of some directions.

\section{Phase transition and stability }

One of the obvious question about black holes is about it's thermodynamical stability. For checking that there are some methods which proves their thermodynamical stability. In some kind of black holes they have phase transition which occur when black hole has critical point. It is interesting to check the thermodynamical stability of our two charged and neutral blackfold. In order to do that we should find some critical points. Generally we know that The critical point occurs when $P$ has an inflection point which is given by applying the below equation on the pressure of two system. One can obtain the following condition,
\begin{equation}
 \label{eq:76}
 \frac{\partial P}{\partial r_{0}}|_{T=T_{c}, r_{0}=r_{c}}=\frac{\partial ^{2} P}{\partial ^{2} r_{0}}|_{T=T_{c}, r_{0}=r_{c}}=0.
 \end{equation}

 We applied the critical condition in eq.(42) for two blackfolds. In that cases we see the results will be same, so one can obtain following following,
 \begin{equation}
 \label{eq:77}
 \frac{\partial P}{\partial r_{0}}|_{T=T_{c}, r_{0}=r_{c}}= -\frac{n\Omega _{n+1} r_{0}^{n-1}}{16\pi G}=0
 \end{equation}

So we do not have any phase transition. The obtained results show that charged and neutral blackfold are thermodynamically stable as we expected.

\section{Conclusion and outlook }

 We calculated the temperature, entropy,chemical potential, charge, pressure, energy density and stress energy tensor for charged and neutral blackfold by  near horizon expansion. It is the familiar method that uses for  spherically symmetric and axisymmetric four dimensional black holes. As the calculation shows, the temperature increase by increasing $n$ and depends inversely by the thickness of the horizon. But charge, entropy, pressure and energy density depend directly to the thickness of horizon. The chemical potential which is properties of charged blackfold depends on $\alpha$ which arises because of electric charge.  The given results confirm the results of other literature.

 As we know the neutral blackfold is Ricci-flat and we reproduced the AdS form of that. Then we calculate the temperature of that AdS blackfold. It proves that the stress tensor is exactly the stress tensor of the Ricci-flat form by the related transformation. Dependency of temperature to $n$ shows the great influence of topology of blackfold to thermodynamical properties.  Finally we checked the phase transition and found that there is not any phase transition and proves both blackfolds are thermodynamically stable. \\

\end{document}